\documentstyle [aps]{revtex}

\begin{document}
\title{Comment on ``The Gowdy $T^3$ Cosmologies Revisited''}
%\\ {\small}}
\author{Beverly K. Berger$^1$, David Garfinkle}
\address{Physics Department, Oakland University, Rochester, MI 48309 USA}
\author{Vincent Moncrief$^1$}
\address{Physics Department, Yale University, New Haven, CT 06520 USA}
\maketitle
\begin{abstract}
A standard and reasonable definition of asymptotic velocity term dominance (AVTD) shows
that the numerical study by Hern and Stewart (gr-qc/9708038) confirms previous results
that generic Gowdy cosmologies on $T^3 \times R$ have an AVTD singularity. 
\end{abstract}
\footnotetext[1]{Visiting
    Scientist: Max-Planck-Institut f\"ur Gravitationsphysik
    (Albert-Einstein-Institut) Schlaatzweg 1, 14473 Potsdam, Germany}

\section*{}
The wave equations for Gowdy universes on $T^3 \times R$ can be written as
\begin{equation}
\label{Peq}
P,_{\tau \tau} - e^{2 P} Q,_{\tau}^2 = e^{-2 \tau} (P,_{\theta \theta} -
e^{2P}Q,_\theta^2)
\end{equation}
and
\begin{equation}
\label{Qeq}
Q,_{\tau \tau} + 2 P,_\tau Q,_\tau = e^{-2\tau}(Q,_{\theta \theta} + 2 P,_\theta
Q,_\theta)
\end{equation}
where $P$ and $Q$ are respectively proportional to the amplitudes of the $+$ and
$\times$ gravitational wave polarizations and $,_a$ signifies ${\partial / {\partial
 a}}$. One usually \cite{IM} defines asymptotic velocity term dominance (AVTD) to mean
that, as
$\tau \to \infty$, $P$ and $Q$ approach a solution to the equations obtained by setting
the right hand sides of (\ref{Peq}) and (\ref{Qeq}) equal to zero. These ``AVTD
equations'' have the exact ``AVTD solution'' \cite{bkbvm}
\begin{eqnarray}
\label{AVTDsoln}
P =& \ln [ \alpha e^{-v \tau} ( 1 + \zeta^2 e^{2 v \tau})] \\
Q =& {{\zeta e^{2 v \tau}}\over {\alpha (1+\zeta^2 e^{2 v \tau})}} + \xi
\end{eqnarray}
with the limiting behavior as $\tau \to \infty$ of 
\begin{equation}
\label{AVTDlim}
P = v \tau, \quad \quad Q = Q_0 
\end{equation}
where $\alpha$,
$\zeta$, and $\xi$ are arbitrary functions of $\theta$, $v \ge 0$ is a function of
$\theta$, and
$Q_0 = 1/\alpha \zeta +
\xi$.  Grubi\u{s}i\'{c} and Moncrief (GM) have argued
\cite{bm1} that, given the limiting AVTD behavior (\ref{AVTDlim}), the spatially
dependent term
$e^{2P}Q,_\theta^2$ in (\ref{Peq}) would ``spoil'' the AVTD solution if $|v| > 1$. Since
$v > 0$ is already required, GM conjecture that, for generic Gowdy models to approach the
AVTD solution as $\tau \to \infty$, one requires $0 \le v < 1$ everywhere in $\theta$
asymptotically except, perhaps, at a set of measure zero. 
If the
Gowdy models have an AVTD singularity, one would expect $P$ and $Q$ to approach
(\ref{AVTDsoln}) and (4) with $0 \le v < 1$ (almost) everywhere for sufficiently large
$\tau$. This behavior has been confirmed in our numerical simulations
\cite{bkbvm,bkb1,bkb2} and has, in fact, been confirmed by HS. To obtain the AVTD
solution for a given system of equations, one sets the spatial derivative terms to zero
by hand. However, our claim on the magnitude of the spatial derivative terms in an AVTD
regime is merely that the dynamics (character of the solution) is that of the AVTD
solution so that the spatial derivative terms do not influence the dynamics. Using this
standard interpretation of an AVTD singularity, one sees that HS's discussion of the
magnitude of spatial derivative terms is irrelevant. In fact, one can use GM's method of
consistency to explain HS's Fig.\/3. This is easy to see if one defines
$\pi_Q = e^{2P}Q,_\tau$. Substitution in the AVTD equations shows that $\pi_Q$ is a
constant (in $\tau$) in the AVTD regime \cite{bm1}. Assume the AVTD limiting form
(\ref{AVTDlim}) for $P$ and $Q$ and substitute these in (\ref{Peq}). One finds
\begin{equation}
\label{PeqGM}
P,_{\tau \tau} = -e^{-2[1-v(\theta)]\tau}[Q'_0(\theta)]^2+e^{-2\tau}v''(\theta) \tau
+\pi_Q^2(\theta) e^{-2 v(\theta) \tau}
\end{equation}
where $' = d/d\theta$.
GM's conjectured restrictions on the magnitude of $v$ mean that either the first or
third term will dominate as $\tau \to \infty$ depending on whether $v$ is less than
or greater than $1/2$. This effect, entirely consistent with the
presumption of an AVTD singularity in the Gowdy models, allows one to reproduce HS's
Fig.\/3 by plotting regions in the $\theta$-$\tau$ plane where $v$ is greater than or
less than $1/2$. Finally, rewriting (\ref{Qeq}) as 
\begin{equation}
\label{piqdot}
\pi_Q,_\tau = e^{-2\tau}(e^{2P}Q,_\theta),_\theta
\end{equation}
shows consistent behavior 
in an AVTD regime since the right hand side becomes
\begin{equation}
\label{rhs}
e^{-2[1-v(\theta)]\tau}[Q_0''(\theta) + 2 v'(\theta) Q_0'(\theta) \tau]
\end{equation}
which decays faster or slower than the third term on the right hand side of
(\ref{PeqGM}) depending on the sign of $v- 1/2$.

HS also criticize previous results \cite{bkbvm,bkb1,bkb2} describing the approach to the
AVTD solution. We have argued \cite{bkb1,bkb2} that the nonlinear terms in (\ref{Peq})
act as potentials
$V_1 = e^{2P} \pi_Q^2$ and $V_2 = e^{-2(\tau - P)}Q,_\theta^2$ to drive the system to
the AVTD regime. GM's conjectured generic AVTD behavior requires (as $\tau \to \infty$)
$P,_\tau > 0$ and $|P,_\tau| < 1$ since $P,_\tau \to v$ asymptotically. As was pointed
out elsewhere \cite{bkb1,bkb2}, if
$P,_\tau < 0$, $V_1$ becomes important eventually causing a bounce which changes the
sign of $P,_\tau$. On the other hand, if $P,_\tau > 1$, $V_2$ will become important
eventually causing a bounce which changes the sign of $P,_\tau -1$. (If this causes
$P,_\tau$ to become negative, there will be another bounce off $V_1$ and so on.) Both
potentials will become negligible when the AVTD limit is reached. Clearly, non-generic
behavior can occur if either $\pi_Q$ or $Q,_\theta$ is precisely zero so that the
effect of the potential is absent. For these coefficients approximately zero, it takes a
long time for the respective bounce to occur since the potential is quite flat. As one
increases the spatial resolution, the simulation grid points are closer to the
non-generic points. Thus the potential is flatter at such points than it would be at
nearby grid points within a coarser resolution. This leads to the effect, as shown in
\cite{bkb1,bkb2}, that it takes longer for a higher resolution simulation to appear to be
AVTD (with $0
\le v < 1$) everywhere, exactly as was found by HS. For example, Fig.\/1 of
\cite{bkb2} displays the maximum value of $v$ on the spatial grid vs $\tau$ for two
spatial resolutions. The coarser resolution was run sufficiently far to yield $v < 1$
everywhere (in constrast to HS's claim that this was not done). The finer resolution
simulation was only run far enough to indicate the expected resolution dependence. This
can also be seen by comparing
$P,_\tau$ vs $\theta$ at a given
$\tau$ for different spatial resolutions. One finds fewer points with $v > 1$ in a
coarser simulation. Careful examination of the waveforms at $\theta$ values where $v >
1$ shows that, at every such point, $Q$ has an extremum in $\theta$ as one expects in
our picture and as has been described in
\cite{bkbvm,bkb1,bkb2}. The non-generic behavior associated with $\pi_Q \approx 0$ leads
to apparent discontinuities in $Q$ as has been discussed elsewhere \cite{bkb1,bkb2}.

In \cite{bkbvm}, we show that, at a fixed value of $\tau$, there exists a spatial
resolution which will resolve any given peak in the waveforms but that, at later $\tau$,
the peak has narrowed so that greater spatial resolution is needed to resolve it. For
example, a peak near $Q,_\theta = 0$ will narrow as $\tau$ increases because ($P,_\tau -
1$) changes sign first at $\theta$ values farthest from the center of the peak. Thus $P$
will continue to grow at the original $v > 1$ only at points ever closer to the center.
This interpretation of our results \cite{bkb1,bkb2} agrees with the fact that the
equations (\ref{PeqGM}) and (\ref{rhs}) obtained by assuming the AVTD solution as $\tau
\to \infty$ are inconsistent if $v(\theta) > 1$ \cite{bm1} unless both $Q'_0$ and
$Q''_0$ vanish. However, one generically expects a set of measure zero at which $Q'_0 =
0$ but $Q''_0 \ne 0$. Thus, using the strict definition of AVTD given by Isenberg and
Moncrief \cite{IM}, one concludes that our numerical results support the view that
generic Gowdy models have a singularity that is AVTD everywhere except, perhaps, at a
set of measure zero. To be consistent with our results, we would expect that an adaptive
mesh refinement (AMR) algorithm would easily observe this narrowing of the spikes and
insert ever more points about the spike center as $\tau$ increases. However, HS do not
find this behavior rather finding that the peaks have ``softened somewhat'' with $v < 1$
everywhere. Since the Lax-Wendroff scheme is known to lead to a decrease in the amplitude
of peaks, we contend that HS's AMR code is not correctly modeling the peaks at late
$\tau$.

Finally, we wish to point out that the Gowdy simultions described in
\cite{bkbvm,bkb1,bkb2} used the symplectic algorithm described in detail in
\cite{bkbvm}. Given sufficient spatial resolution, this algorithm is stable for
evolution of the Gowdy equations. (In \cite{bkbvm}, the resultant data were smoothed
prior to graphing in some of the figures. This ``averaging'' had nothing to do with the
simulation algorithm.)

In summary, the AMR algorithm of HS has generated a Gowdy evolution
that confirms previous analytic \cite{bm1} and numerical \cite{bkbvm,bkb1,bkb2} results
except close to the centers of peaks where we believe our algorithm to be more acurate.

\section*{Acknowledgements}
B.K.B.~and V.M.~would like to thank the Albert Einstein Institute at Potsdam for
hospitality. This work was supported in part by National Science Foundation
Grants PHY9507313, PHY9722039, and PHY9503133.

\end{document}